\RequirePackage{lineno}
\documentclass[rmp,twocolumn,floatfix]{revtex4}
\usepackage[english]{babel}
\usepackage[utf8x]{inputenc}
\usepackage[T1]{fontenc}
\usepackage{amsmath,amsfonts,amssymb,eucal,eurosym,textcomp}
\usepackage{color}
\usepackage{graphicx}

\graphicspath{{./figures/}}
\definecolor{drab}{rgb}{0.59, 0.44, 0.09}

\newcommand{\figurewidth}{\linewidth}
\newcommand{\figurewide}{0.99}
\newcommand{\figurenarrow}{0.7}
\newcommand{\protein}[1]{#1}
\newcommand{\gene}[1]{\textit{#1}}
\newcommand{\gag}{\gene{gag}}
\newcommand{\env}{\gene{env}}
\newcommand{\rev}{\gene{rev}}
\newcommand{\pol}{\gene{pol}}
\newcommand{\nef}{\gene{nef}}

\newcommand{\FIG}[1]{\figurename~\ref{fig:#1}}
\newcommand{\TAB}[1]{\tablename~\ref{tab:#1}}
\newcommand{\MM}[1]{Materials and Methods}
\newcommand{\SNPlong}[1]{single nucleotide polymorphism}
\newcommand{\SNP}[1]{SNP}
\newcommand{\rTM}{$^\circledR$}
\newcommand{\C}{$^\circ$C}

\usepackage{hyperref}
\begin{document}
\title{Population genomics of intrapatient HIV-1 evolution}

\author{Fabio Zanini$^{1}$, Johanna Brodin$^{2}$, Lina Thebo$^{2}$, Christa Lanz$^{1}$, G\"oran Bratt$^4$, Jan Albert$^{2,3*}$ and Richard A.~Neher$^{1*}$}
\affiliation{\scriptsize $^{1}$\mbox{Evolutionary Dynamics} and Biophysics, Max Planck Institute for Developmental Biology,\\ 72076 T\"ubingen, Germany \\
$^{2}$Department of Microbiology, Tumor and Cell Biology, Karolinska Institute, Stockholm, Sweden\\
$^3$Department of Clinical Microbiology, Karolinska University Hospital, Stockholm, Sweden\\
$^4$Department of Clinical Science and Education, Venhälsan, Stockholm South General Hospital, Stockholm, Sweden\\
$^{*}$ corresponding author.
}
\date{\today}
\begin{abstract}
Many microbial populations rapidly adapt to changing environments with multiple variants competing for survival. To quantify such complex evolutionary dynamics in vivo, time resolved and genome wide data including rare variants are essential. We performed whole-genome deep sequencing of HIV-1 populations in 9 untreated patients, with 6-12 longitudinal samples per patient spanning 5-8 years of infection. 
We show that patterns of minor diversity are reproducible between patients and mirror global HIV-1 diversity, suggesting a universal landscape of fitness costs that control diversity. Reversions towards the ancestral HIV-1 sequence are observed throughout infection and account for almost one third of all sequence changes. Reversion rates depend strongly on conservation. Frequent recombination limits linkage disequilibrium to about 100bp in most of the genome, but strong hitch-hiking due to short range linkage limits diversity.
\end{abstract}
\maketitle


The human immunodeficiency virus 1 (HIV-1) is a paradigmatic example of a rapidly adapting population characterized by high diversity, strong selection, and recombination. HIV-1 has originated from multiple zoonotic transmissions from apes in the early part of the 20th century \cite{sharp_origins_2011}, one of which gave rise to the worldwide pandemic. This lineage is called group M and has diversified into different subtypes at a rate of about 1 in 1000 substitutions per year \cite{lemey_HIV_2005,li_integrated_2015}. This diversification has been extensively characterized and sequences of tens of thousand of HIV-1 variants are available in the Los Alamos National Laboratories HIV database (LANL) \cite{foley_hiv_2013}.

The evolution of HIV-1 ultimately takes place within infected individuals and can be observed directly in longitudinal samples of virus populations from the same individual. The detailed knowledge of HIV-1 biology paired with historical samples make HIV-1 an ideal system to study general features of evolution at high mutation rates and strong selection that are otherwise only accessible in evolution experiments \cite{miralles_clonal_1999,elena_evolution_2003}.

Longitudinal data has been used to characterize escape from cytotoxic T-cells (CTL) \cite{salazar-gonzalez_genetic_2009,kearney_human_2009,liu_dynamics_2011}, track evolution driven by the humoral immune response against HIV-1 \cite{richman_rapid_2003,shankarappa_consistent_1999} and to study the emergence of drug resistance \cite{paredes_vivo_2009}. In a pioneering study, \citet{shankarappa_consistent_1999} characterized HIV-1 evolution of parts of the \protein{gp120}{} envelope protein over approximately 6-12 years in 11 patients, demonstrating consistent evolution of diversity and divergence. While this early study was limited to about 10 sequences per sample, next-generation sequencing technology today allows deep characterization of intrapatient HIV-1 variability including rare mutations \cite{hedskog_dynamics_2010, fischer_transmission_2010, tsibris_quantitative_2009}. However, these studies had limited follow-up (i.e.\ only one or a few time points were sequenced) or focused on small parts of the genome. \citet{henn_whole_2012} was the first study to sequence the majority of the virus genome, but is restricted to a single patient with limited follow-up.

To develop a comprehensive and quantitative understanding of the evolution and diversification of HIV-1 populations, we generated a whole-genome, deep-sequencing data set covering 9 patients over 5 to 8 years with 6 to 12 time points per patient. Importantly, the data set covers the entire genome such that no substitution is missed and includes early samples defining the initial population. To our knowledge, this is the only whole-genome deep sequencing data set with long follow-up of multiple patients. 

Below, we first describe the methodology we developed to sequence the entire HIV-1 genome at a great depth while preserving linkage over several hundred base pairs. We then analyze the intrapatient evolution of HIV-1 and show that the minor variants in the virus population explore sequence space in predictable fashion at the single site level. At the same time, we observe a strong tendency for reversion towards the global HIV-1 consensus. Reversion is faster at sites that are more conserved at the global level -- suggesting a direct relationship between intrapatient fitness cost and global conservation. We further find frequent recombination, which allows the viral population to evolve independently in different regions of the genome. Nevertheless, recombination is not frequent enough to decouple mutations closer than 100 base pairs and we observe signatures of hitch-hiking at short distances \cite{smith_hitch-hiking_1974}.


\section*{\large{Results}}
The study included nine HIV-1-infected patients who were diagnosed in Sweden between 1990 and 2003. Data from two additional patients were excluded from analysis because of suspected superinfection and failure to amplify early samples with low virus levels, respectively. The patients were selected to have a relatively well-defined time of infection and have been treatment-naive for a minimum of five years. Patients diagnosed in recent years rarely fulfill these inclusion criteria because therapy is almost universally recommended, but this was not the case when the study patients were diagnosed. Basic characteristics of the patients and the samples are presented in \TAB{patients}. The method to estimate the time since infection (ETI) is described in \MM{}. Of the nine patients, eight were males, seven were men who have sex with men (MSM), two were heterosexually infected (HET), and eight were infected in Sweden. We retrieved longitudinal biobank plasma samples covering the time period during which the patients had been untreated (5.5-8.3 years). The number of samples per patient ranged between 6 and 12 and in total we investigated 73 samples. The median plasma HIV-1 RNA level was 12,000 copies/ml for the samples with available data on RNA levels. Some samples had low RNA levels, which partly is explained by the study design that required that the patients were treatment-naive for at least five years. For detailed information about each individual sample see supplementary table S1 (\texttt{S1\_samples.csv}). 

\begin{table*}[t]
\begin{center}
\scriptsize
\begin{tabular}{|l|c|c|c|c|c|c|c|c|c|c|c|c|}
\hline
Patient & Gender  & Transmission & Subtype & Age* & Fiebig & BED* & No. of & First sample & Last sample & \multicolumn{3}{c|}{HLA type} \\
	&   & route &      &  [years] & stage* & [ODn] & samples     & [days] & [years] &   A & B & C            \\ \hline
 p1	& F & HET & 01\_AE& 37& IV &  0.41 & 12 &  49 & 8.0 & 02/02 & 08/15 & 03/06\\ \hline
 p2 & M & MSM &      B & 32 &  V &  0.17 &  6 &  74 & 5.5 & 01/24 & 08/39 & 07/12\\ \hline
 p3 & M & MSM &      B & 52 & VI &  0.89 & 10 & 104 & 8.3 & 02/11 & 15/44 & 03/16\\ \hline
 p5 & M & MSM &      B & 38&III-IV& n.a. &  7 & 132 & 5.9 & 03/33 & 14/58 & 03/08\\ \hline
 p6 & M & HET  &      C & 31 & IV &  0.29 &  7 &  46 & 7.0 & 02/02 & 44/51 & 05/16\\ \hline
 p8 & M & MSM &      B & 35 &  V &  0.15 &  7 &  64 & 6.0 & 03/32 & 07/40 & 02/07\\ \hline
 p9 & M & MSM &      B & 32 & VI &  0.27 &  8 & 106 & 8.1 & 25/32 & 07/44 & 04/07\\ \hline
p10 & M & MSM &      B & 34 & II &  0.10 &  9 &  18 & 6.1 & 32/32 & 44/50 & 06/16\\ \hline
p11 & M & MSM &      B & 53 & VI &  1.22 &  7 & 167 & 5.5 & 02/32 & 39/44 & 05/12\\ \hline
\end{tabular}
\caption{Summary of patient characteristics. Sample times from estimated date of infection. $^{*}$, at base line; MSM, men who have sex with men; HET, heterosexual.}
\label{tab:patients}
\end{center}
\end{table*}

\subsection*{HIV-1 whole-genome deep sequencing}
\begin{figure}[!ht]
\begin{center}
\includegraphics[width=\figurenarrow\linewidth]{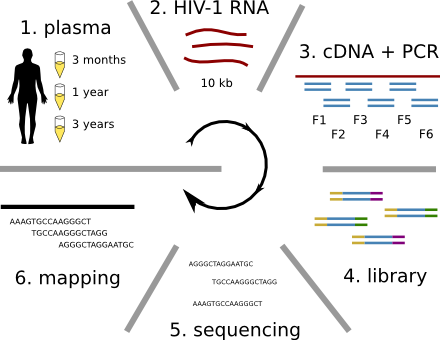}
\includegraphics[width=\figurenarrow\columnwidth]{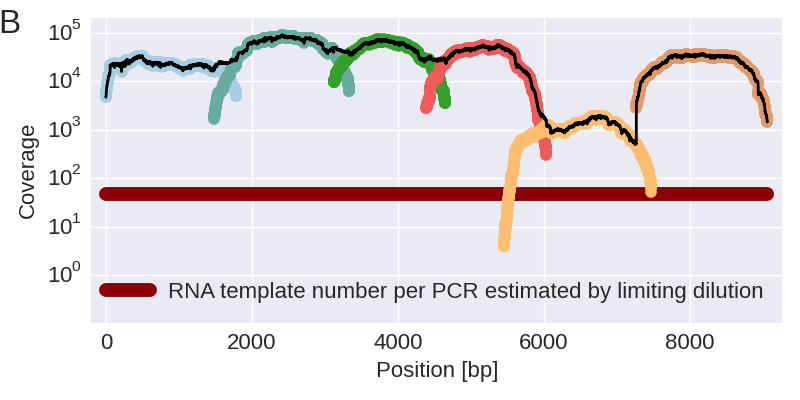}\\[3ex]
\includegraphics[width=\figurenarrow\linewidth]{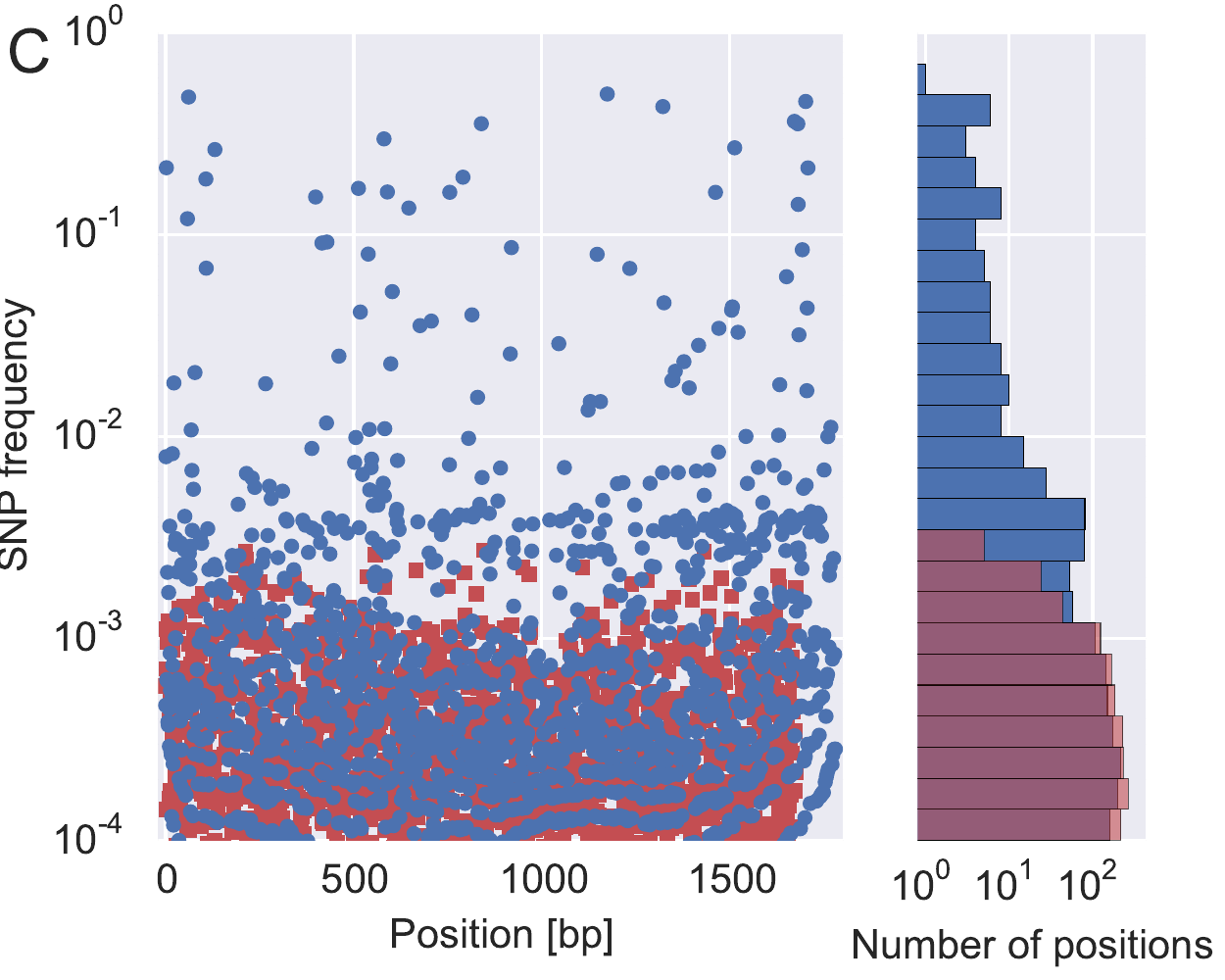}
\caption{{\bf Sequencing, coverage, and error rates}. (A) schematics of the sample preparation protocol, see text and \MM{} for details. (B) Read coverage for a representative sample. Coverage of separate PCR amplicons is shown in different hues, the black line is the total coverage. The coverage of PCR fragment F5 is lower than the other amplicons, but it is still larger than number of input HIV-1 RNA molecules; this situation is typical in our samples. (C) Each blue circle corresponds to a \SNP{} frequency in amplicon F1 of a late sample of patient 11, while red squares are \SNP{} frequencies in the sequence data generated from 10,000 copies of plasmid NL4-3. The histogram on the right shows the distribution of \SNP{} frequencies in the patient sample and the control. Minor \SNP{}s observed in reads generated from the plasmid, which represent PCR and sequencing errors, did not exceed 0.3\%.
}
\label{fig:sketch}
\label{fig:afEx}
\label{fig:coverage}
\end{center}
\end{figure}
We extracted viral RNA and amplified and sequenced it as explained in \MM{}. The basic steps are illustrated in \FIG{afEx}. Viral RNA was extracted from 400~µl patient plasma and used for one-step RT-PCR amplification with six overlapping primer sets that span almost the complete HIV-1 genome, similar to the strategy developed by \citet{gall_universal_2012}. Sequencing libraries were made starting from 0.1 -- 1.5 ng of DNA with a stringent size selection for long inserts (\textgreater 400 bp); the sequencing reads were filtered and mapped to the individual fragments defined by the PCR primers. Sequencing was performed on the Illumina MiSeq platform and sequence reads were quality filtered and assembled using an in-house data processing pipeline (see \MM{}). In total approximately 100 million reads passed the quality filtering. The coverage varied considerably between samples and amplicons, but was mostly of the order of several thousands or more, see \FIG{coverage}B.

Importantly, sequencing depth is determined not only by coverage but also by template availability and sequencing errors. We performed a number of control experiments to quantify templates and assess the accuracy of estimates of frequencies of \SNP{}. The results are summarized in the following section and described in detail in \MM{}.

We quantified the number of HIV-1 genomes that contributed to each sequencing library by PCR limiting dilution. Comparison with routine plasma HIV-1 RNA level measurements performed at time of sampling showed that the median template cDNA recovery efficiency was 30\%. Most samples had low to moderate HIV-1 RNA levels and hence template availability, rather than coverage, determined the sequencing depth, see \FIG{coverage}B.

We estimated the error rate of the PCR and sequencing pipeline by amplifying and sequencing a plasmid clone. \FIG{afEx}C compares the \SNP{} frequencies observed in a clone to those observed in a patient sample. After quality filtering, PCR and sequencing errors never exceeded 0.3\% of reads covering a particular position. To detect and control for variation in PCR efficiency among fragments and skewed amplification of different variants, we compared frequencies of variants in overlaps between the six amplicons. A \SNP{} in the overlap is amplified and sequenced twice independently and the concordance of the two measurements of variant frequencies was used to estimate the fragment specific depth, see \MM{} and \FIG{templateQuant}. Frequency estimates were often reproducible to within 1\%. Sometimes, however, variant recovery was poor (mostly in fragment 5) and frequency estimates less accurate. 

We minimized PCR recombination by reducing the number of PCR cycles and optimizing the reaction protocol (see \MM{}, \citet{mild_performance_2011,di_giallonardo_next-generation_2013}). Control experiments using mixtures of two cultured virus populations show that less than 10\% of reads have experienced RT-PCR recombination.

Taken together, our control experiments show that depending on the sample and fragment, we could estimate frequencies of SNPs down to 1\% accuracy (corresponding to several thousand effective templates). In some cases, however, the template number was low or template recovery poor such that only presence or absence of a variant could be called. Furthermore, \SNP{}s remained linked through PCR and sequencing.


\begin{figure}[ht]
\begin{center}
\includegraphics[width=\figurewide\figurewidth]{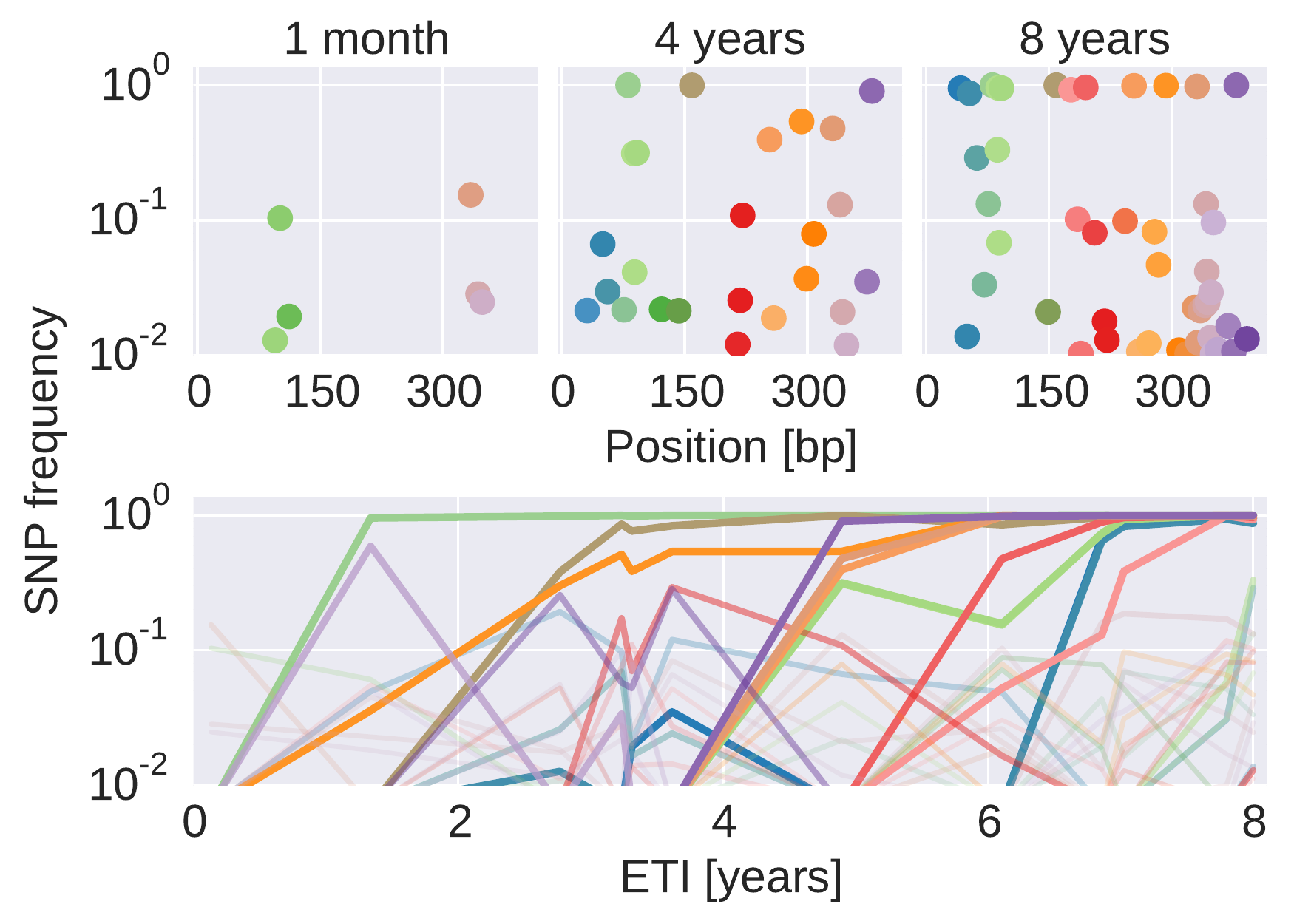}
\caption{{\bf The dynamics of \SNP{} frequencies.} The upper panels show \SNPlong{} frequencies along \protein{p17} at three time points in patient p1. The lower panel shows the trajectories of \SNP{}s through time. Color corresponds to position in the sequence. Trajectory that reach high frequencies are shown with thicker and more opaque lines. Analogous data is available for all patients for most of the HIV-1 genome.
}
\label{fig:SNVexample}
\end{center}
\end{figure}

\subsection*{Consistent evolution across the entire genome}
In most patients, the virus populations was initially homogeneous and diversified over the years, as expected for an infection with a single or small number of similar founder viruses \cite{keele_identification_2008}. In two patients, p3 and p10, the first sample displayed diversity consistent with the transmission of several variants from the same donor. 
For each of the nine patients we reconstructed the HIV-1 genome sequence of the first sample by an iterative mapping procedure described in \MM{}. This initial consensus sequence approximates the sequence of the founder virus(es).

\FIG{SNVexample} shows an example of the dynamics of frequencies of \SNP{} relative to the founder sequence over time, where each dot (top) or line (bottom) represents the frequency of a nucleotide different from the founder sequence. Interactive versions of this graph are available for the entire genome of all patients at \url{hiv.tuebingen.mpg.de}.

\begin{figure}
\begin{center}
\includegraphics[width=\figurewide\figurewidth]{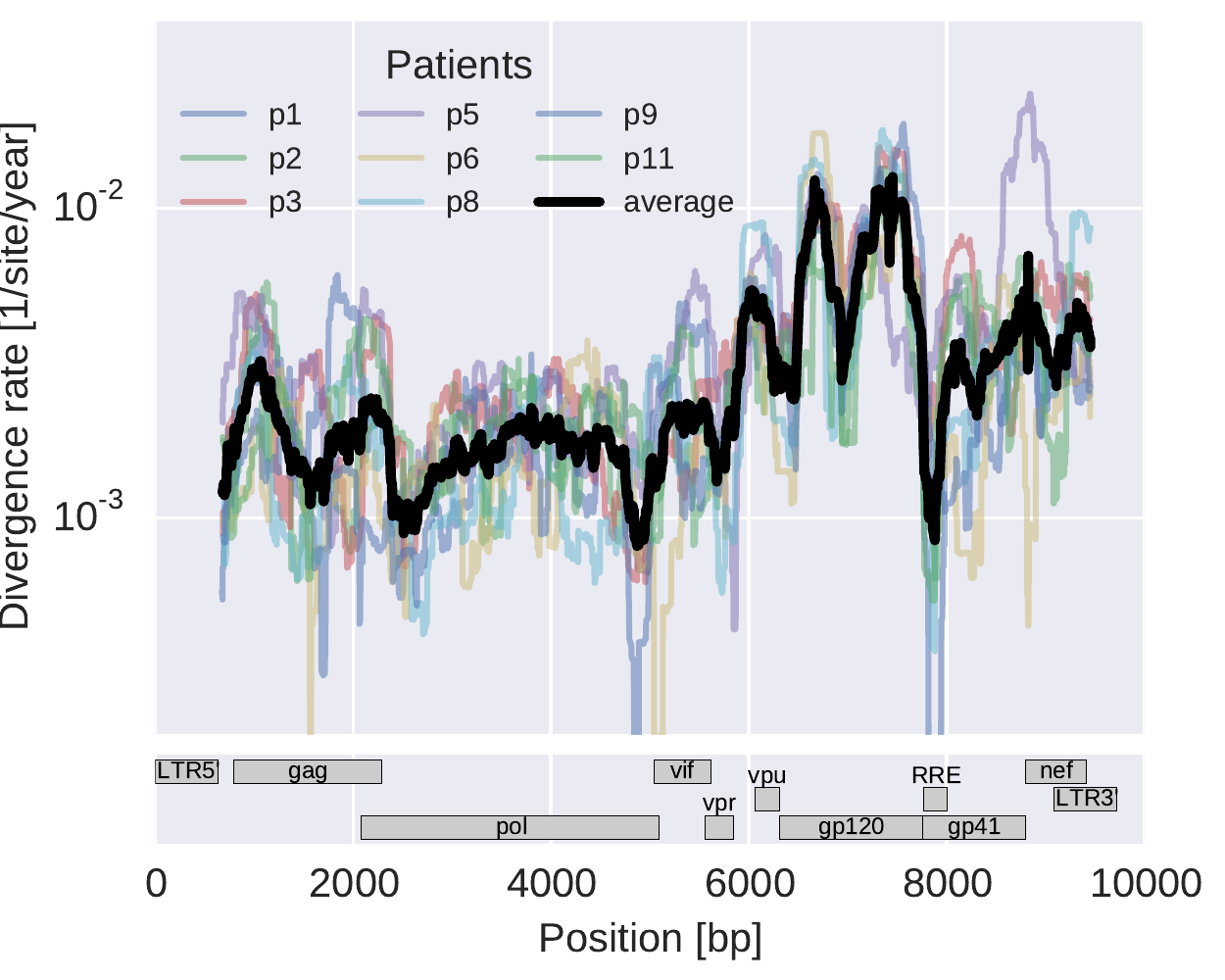}
\caption{{\bf Consistent evolution across the viral genome}. The figure shows the rate of sequence divergence averaged in a sliding window of length 300 bp for individual study participants (in color) and averaged over all (black).
Rapidly evolving (V loops in \protein{gp120}) and conserved (RRE) regions are readily apparent. The divergence rates are determined by linear regression of the distance from the putative founder sequence against ETI. This distance includes contributions of minor variants.  All positions are given in HxB2 numbering.
}
\label{fig:subsRates}
\end{center}
\end{figure}

To measure the rate at which the virus population accumulates mutations, we calculated the average distance of each sample from the founder sequence in 300bp windows. Regressing this divergence against time yields the rate of divergence in different regions of the genome, see \FIG{subsRates}. As expected, some regions such as the variable loops in \protein{gp120} and \nef{} evolve faster, while enzymes -- protease (PR), reverse transcriptase (RT) -- and the \rev{} response element (RRE) evolve more slowly. The rate of divergence varies by about a factor of 10 along the genome, but it is consistent with typically about 1.5-fold differences across patients (standard deviation of log2(fold change) $0.6 \pm 0.2$).
The overall pattern of the rate of mutation accumulation agrees with a recent map of HIV genome-wide variation from a population perspective \cite{li_integrated_2015} and correlates well with entropy in a large HIV-1 group M alignment (Spearman's $\rho=0.7$ after the same smoothing).

\subsection*{Minor variants reproducibly explore global HIV-1 diversity}
Having found that coarse patterns of divergence are comparable among patients, we asked whether intrapatient diversity at individual sites in the viral genome follows general and predictable patterns. To this end, we compared diversity at each position to the diversity observed in HIV-1 group M (see \MM{}).

\begin{figure}
\begin{center}
\includegraphics[width=\figurewide\figurewidth]{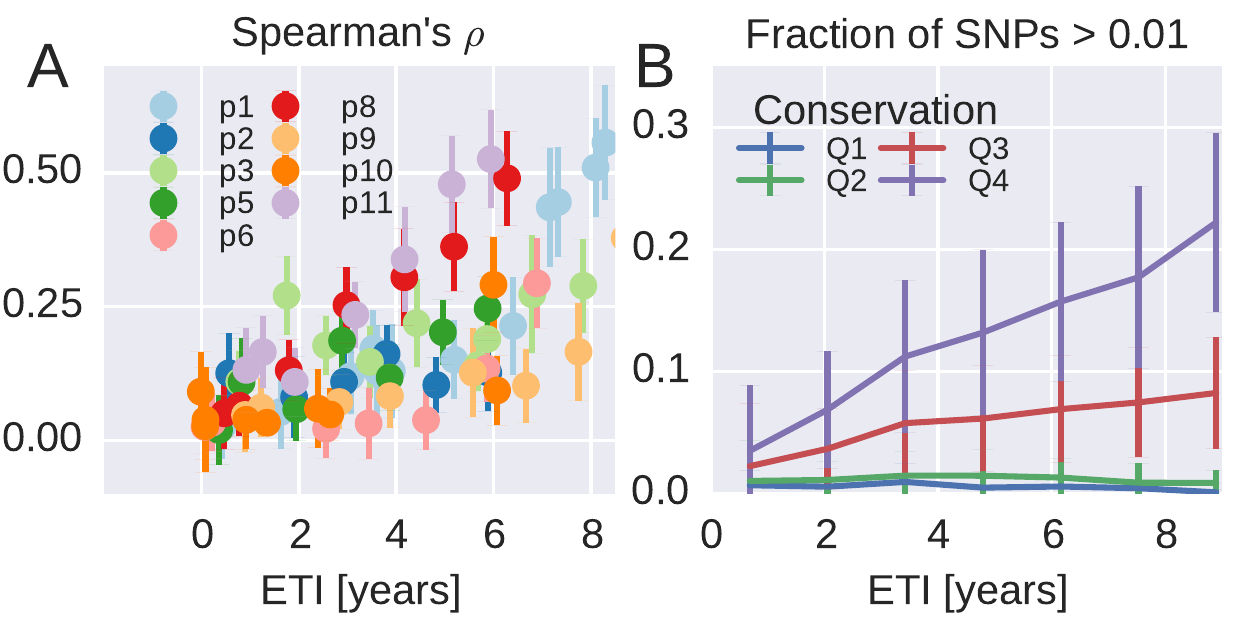}
\caption{{\bf Within patient variation mirrors global variation}. (A) Intrapatient variation at individual sites is correlated with diversity at homologous positions in an alignment of sequences representative of HIV-1 group M. This correlation increases reproducibly throughout the infection. Error bars show standard deviations over genomic regions. (B) Similarly, the fraction of sites with minor variants above 1\% increases over time at the least constrained positions (quartiles Q3 and Q4), while few sites in the most conserved quartiles (Q1 and Q2) are polymorphic.
}
\label{fig:corrPatSub}
\end{center}
\end{figure}

\FIG{corrPatSub}A shows the rank correlation between the site-by-site diversity in each patient and a global collection of HIV-1 sequences, both measured by Shannon entropy (see \MM{}). In all cases correlation with cross-sectional diversity was initially low, as expected for largely homogeneous populations. As diversity increases within patients, it tends to accumulate at positions that are not conserved, resulting in a rank correlation of about 0.4-0.5 after about 8 years. These correlations are individually significant and reproducible among different genomic regions (error bars in \FIG{corrPatSub}).

\FIG{corrPatSub}B offers an alternative perspective on the exploration of sequence space by the HIV-1 populations. We classified nucleotide positions in the genome into four categories ranging from highly conserved positions to less conserved positions within group M (Q1 -  Q4) using the same alignment as above. Next we asked what fraction of sites within these categories show intrapatient variation at a level of at least 1\%. For the least conserved positions, this fraction increased steadily to about 20\% after 8 years, while less than 1\% of the most conserved sites shows variation at the 1\% level. This latter fraction rapidly saturates and does not increase over time. Since variant amplification, sequencing and variant calling does not use any information on cross-sectional conservations, the absence of variation above 1\% at conserved sites is further evidence that our amplification and sequencing pipeline does not generate spurious variation. Other thresholds yield similar results.

Taken together, the observations in \FIG{corrPatSub} show that variation is not limited by the mutational input, but that HIV-1 populations accumulate diversity wherever mutations do not severely compromise virus replication. At the single nucleotide level, the spectrum of mutational possibilities is explored reproducibly and the level of within-host diversity is predicted by time since infection and cross-sectional diversity. Conserved positions are typically monomorphic even in deep samples.

\subsection*{The majority of nonsynonymous substitutions are positively selected}

\begin{figure}
\begin{center}
\includegraphics[width=\figurewide\figurewidth]{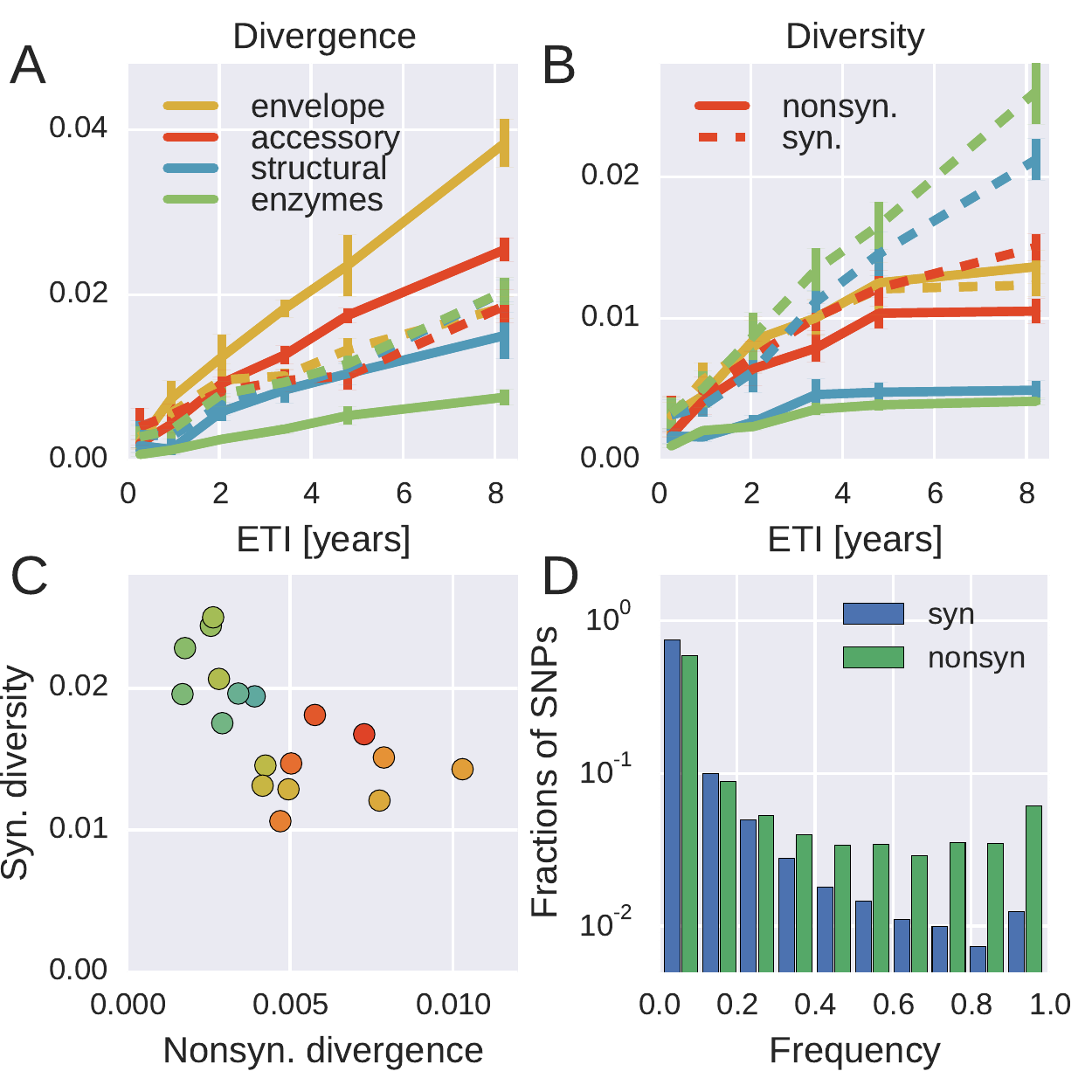}
\caption{{\bf Distinct patterns of evolution across mutation types and regions.} Panel A shows divergence at nonsynonymous (solid) and synonymous (dashed) positions over time for different genomic regions averaged over all patients. While synonymous divergence is very similar in different regions, nonsynonynous divergence varies. Panel B shows diversity though time. Regions with high nonsynonymous diversity (and divergence) tend to have low synonymous diversity. Error bars represent standard deviations of patient bootstrap replicates. Panel C shows the anti-correlation between the rate of nonsynonymous divergence and synonymous diversity in 1kb windows across the genome (color indicates position on the genome blue$\to$green$\to$yellow$\to$red). Panel D shows the site frequency spectrum of synonymous (blue) and nonsynonymous (green) \SNP{}s.
}
\label{fig:divergenceConsPop}
\label{fig:sfsSynNonsyn}
\end{center}
\end{figure}
In addition to the reproducible patterns of minor diversity, virus evolution is characterized by adaptations that are specific to the host. Immune selection results in escape mutations that rapidly spread through the population \cite{walker_T-cell_2012,richman_rapid_2003}. Such mutations tend to be nonsynonymous, i.e., change the viral proteins, while evolution at synonymous sites is expected to be conservative. Nevertheless, synonymous mutations are affected by ``selective sweeps'' of linked nonsynonymous mutations \cite{smith_hitch-hiking_1974}.

To quantify the degree at which the evolutionary dynamics of HIV-1 is dominated by selective sweeps, we calculated divergence and diversity separately nonsynonymous and synonymous in different parts of the HIV-1 genome. \FIG{divergenceConsPop}A compares nonsynonymous (solid lines) and synonymous divergence (dashed lines) in different regions of the genome. In agreement with the results presented in \FIG{subsRates}, the observed rate of evolution at nonsynonymous sites differed substantially between genomic regions, with \env{} being fastest and \pol{} slowest. Divergence at synonymous sites, however, varies very little between different genomic regions indicating random accumulation of synonymous mutations (rather than positive selection).

\FIG{divergenceConsPop}B shows the corresponding plot for diversity, i.e., the distances between sequences from the same sample. Diversity at nonsynonymous sites (solid lines) saturates after about 2-4 years, suggesting that nonsynonymous \SNP{}s either stay at low frequency because they are deleterious, or rapidly increase in frequency and fix without contributing much to diversity. Synonymous diversity increases steadily in the 5' part of the genome (structural and enzymes), while it saturates in the 3' half of the genome after a few years -- the exact opposite of nonsynonymous divergence.

Indeed, we observe a strong anti-correlation between synonymous diversity and nonsynonymous divergence, which is further quantified in \FIG{divergenceConsPop}C. This suggests that frequent non-synonymous substitutions limit synonymous because they drive  linked synonymous mutations to fixation or to extinction \cite{smith_hitch-hiking_1974}. We will quantify linkage and recombination below, but the differences in diversity accumulation already suggest that linkage is restricted to short distances.

The contrasting behavior of synonymous and nonsynonymous \SNP{}s is also seen in the \SNP{} frequency spectrum -- the histogram of \SNP{} abundance -- shown in \FIG{sfsSynNonsyn}D. While the spectra agree for frequencies below 20\%, synonymous mutations are strongly underrepresented at higher frequencies (Fisher's exact test at frequency 0.5, $p$-value $\sim 10^{-80}$). This corroborates the interpretation that due to substantial recombination, sweeping nonsynonymous mutations only occasionally ``drag'' adjacent synonymous mutations to fixation. Synonymous mutations rarely rise in frequency because of their own effect on fitness, since they usually have small or deleterious phenotypic effects and do not contribute directly in immune evasion \cite{zanini_quantifying_2013}. The about 5-fold excess of nonsynonymous over synonymous \SNP{}s at high frequencies (see \FIG{sfsSynNonsyn}D) shows that the majority of common nonsynonymous mutations spread due to positive selection.

Next, we sought to quantify what fraction of nonsynonymous divergence is driven by escape from cytotoxic T-lymphocytes (CTLs). Four-digit HLA types were determined for all patients and a set of putatively targeted HIV-1 epitopes were determined using the epitope binding prediction tool MHCi (\url{tools.immuneepitope.org/mhci}).
We then asked whether we observed more nonsynonymous substitutions in epitopes predicted to be targeted than expected by chance (excluding the variable loops of \protein{gp120} and the external part of \protein{gp41}, see \MM{}). We found a significant enrichment by a factor 1.9 in the putatively targeted region ($p$-value $<3 \times 10^{-6}$), corresponding to 5.5 excess nonsynonymous substitutions, whereby the total number of nonsynonymous substitutions per patient is on average 43 (median, quartiles 36 -- 64). Note that the set of predicted epitopes will contain false positives and lack true epitopes, hence the actual number of CTL driven substitutions could be higher as for example suggested by \citet{allen_selective_2005} who report that roughly half of non-envelope mutations are associated with CTL responses.

\begin{figure}
\begin{center}
\includegraphics[width=\figurewide\figurewidth]{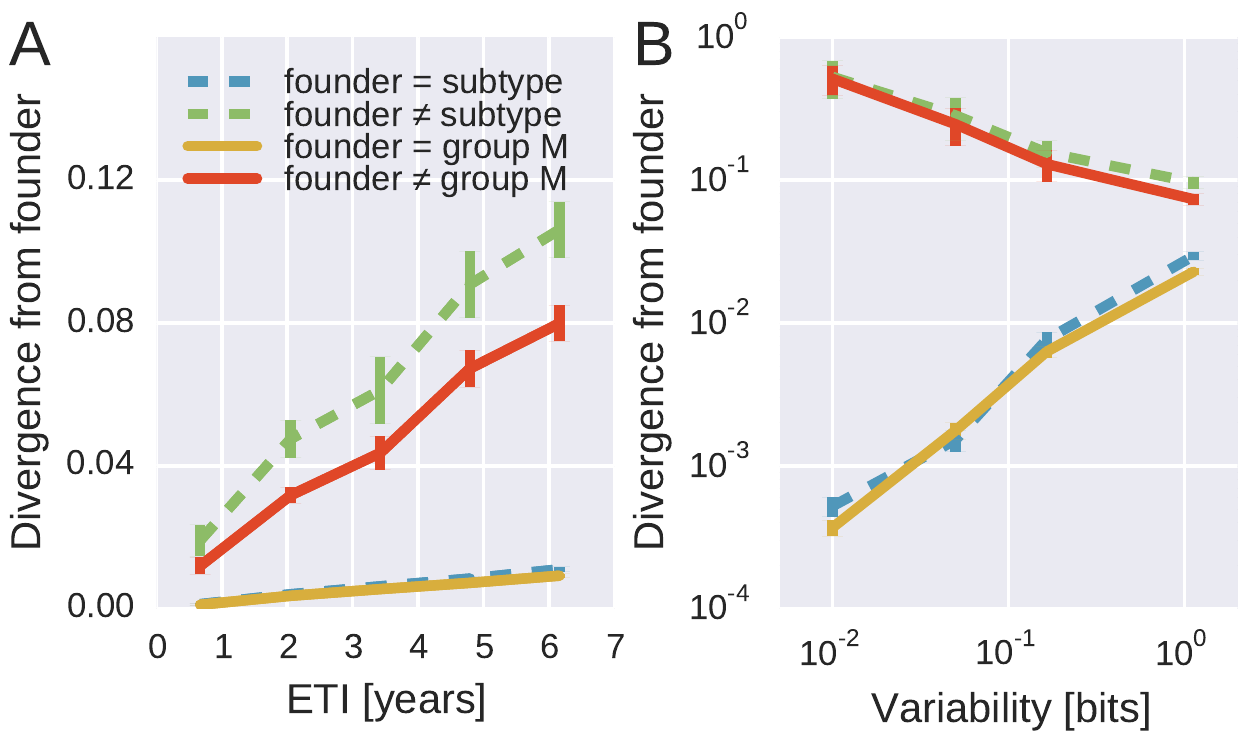}
\caption{{\bf Rapid reversion at conserved sites}. Panel A: Sites where the founder sequence differed from the subtype or group M consensus (upper curves) diverged about 10 fold more rapidly than sites that initially agreed with the consensus (lower curves). Panel B: The rate of reversion increased with conservation (lower variability), while divergence away from consensus showed the opposite behavior (divergence is measured at 5-6 years). Error bars report the standard deviation of patient bootstraps.}
\label{fig:SFSawayto}
\end{center}
\end{figure}

\subsection*{Extensive reversion towards consensus}
Many CTL escape mutations reduce the replicative capacity of the virus and it is known that such escape mutations often revert upon transmission to a host in which the corresponding epitope is not targeted \cite{leslie_hiv_2004,friedrich_reversion_2004,herbeck_human_2006}. The balance between escape and reversion results in association between specific escape mutations and the HLA types of the hosts \cite{kawashima_adaptation_2009,palmer_integrating_2013}. In a large population of hosts, the most common allele at a specific site is likely favored, while rare alleles tend to be escape mutations that reduce viral replicative capacity \cite{carlson_selection_2014}.

To quantify patterns of reversion and fitness cost, we classified sites in the approximate founder sequence of the viral populations in each subject as being identical or different from the HIV-1 group M consensus. \FIG{SFSawayto}A shows the fraction of sites where the founder nucleotide is replaced by a mutant during the infection. This fraction is about 10 fold higher if the founder nucleotide differs from the group M consensus than if it is identical to the group M consensus. Reversion towards group M consensus occurs at a roughly constant rate throughout the observation time (5-8 years).

Of all changes accumulated by the viral populations, $30\pm 2.5$\% are reversions towards group M consensus (mean and standard deviation of patient bootstraps after 4-7 years). Similar results are found when comparing to the subtype consensus of each patient virus ($24\pm 2.5\%$). Reversions are between 4 and 5 times more frequent than expected in absence of a reversion bias (7.8\% and 4.5\%, respectively). These findings agree with results by \citet{allen_selective_2005}, who report that about 20\% of amino acid substitutions are reversions.

By focusing on sites where the founder virus differed from the group M consensus, we are predominantly looking at weakly conserved sites. To control for conservation, we carried out the same analysis after stratifying sites by overall level of conservation. \FIG{SFSawayto}B shows the result of this analysis, focusing on samples after 5-6 years for the sake of clarity. We find that the rate of reversion is highest at the most conserved sites. Almost 50\% of all non-consensus positions at highly conserved sites had reverted to consensus after about 5 years -- an almost 1000-fold excess. Even at the least conserved sites, divergence towards group M consensus exceeded divergence away from group M consensus by a factor of 3. These results suggest that the global HIV-1 group M consensus sequence represents an ``optimal'' HIV-1 sequence, which acts as an attractor for the evolutionary dynamics within hosts. This attraction is strongest at conserved sites, but extends to the least conserved sites in the genome.

\subsection*{Lack of long-range linkage due to frequent recombination}
\begin{figure}
\begin{center}
\includegraphics[width=\figurewide\figurewidth]{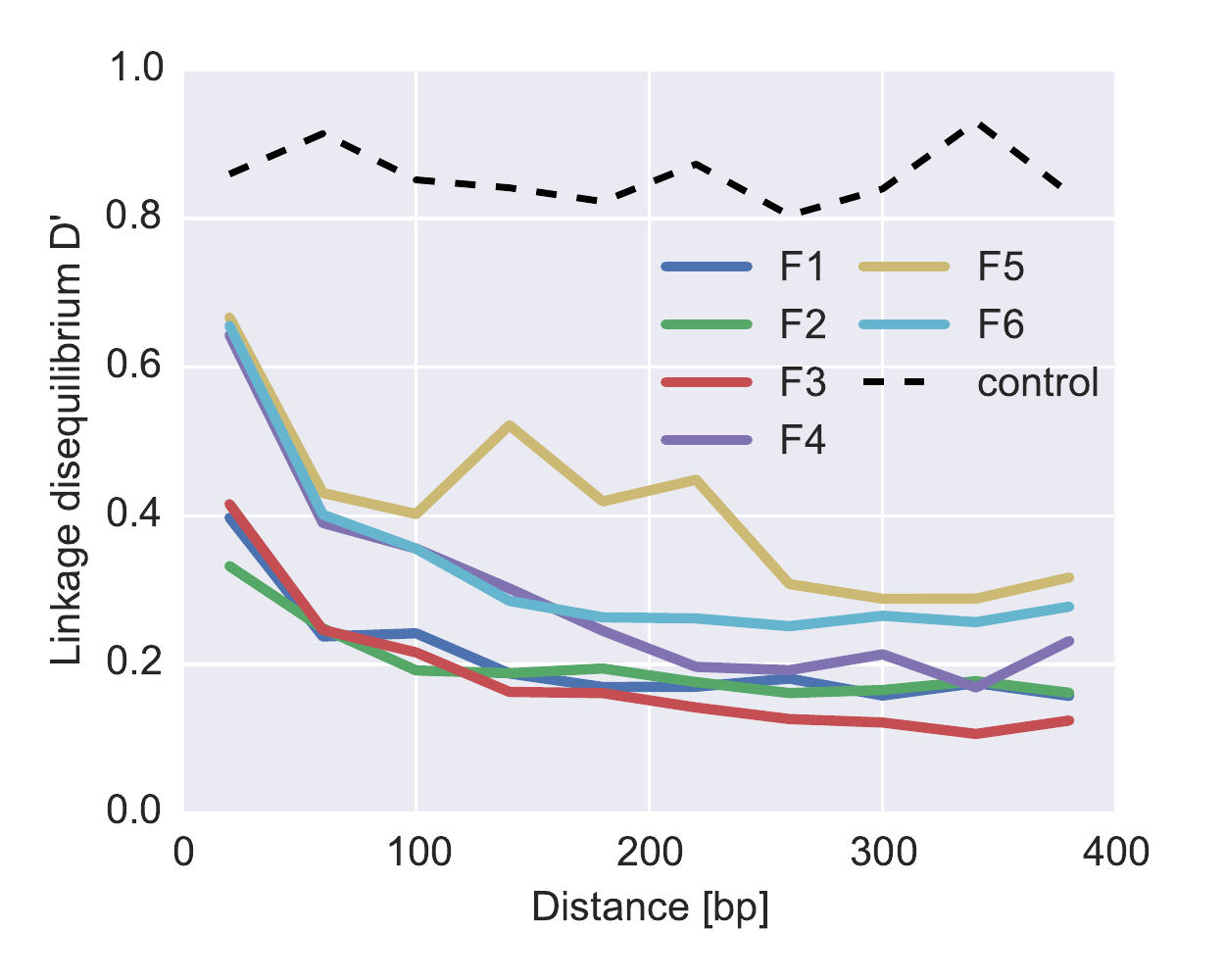}
\caption{{\bf Linkage and recombination. } Linkage disequilibrium decays rapidly with distance between \SNP{}s. Colored lines correspond to the different fragments, each averaged over patients. The dashed line shows data from a control experiment for PCR recombination, where two cultured virus populations were mixed. No PCR recombination is observed.}
\label{fig:LD}
\end{center}
\end{figure}

To quantify the decoupling of \SNP{}s by recombination, we calculated linkage disequilibrium (LD) between \SNP{}s as a function of distance for each of the six fragments, see \FIG{LD}. For most fragments, we observed a consistent decrease of LD over the first 100-200 bps, with fragment 5 being an exception with linkage of mutations at longer distances. Importantly, our linkage control (a 50/50 mix of two distinct virus isolates and a total of 1,250 RNA molecules per PCR fragment) shows no decay of LD with distance at all, suggesting negligible RT-PCR recombination.

The observed decay of LD in patient samples is consistent with a recombination rate of $10^{-5}/bp/day$ as estimated in \cite{neher_recombination_2010,batorsky_estimate_2011}. Our reasoning proceeds as follows. \FIG{SFSawayto}B indicates that diversity accumulates over a time frame of 2-4 years, i.e., about 1,000 days. Recombination at a rate of $10^{-5}/bp/day$ hits a genome on average every 100 bps in 1000 days. Mutations further apart than 100bps are hence often separated by recombination and retain little linkage consistent with the observed decay length in \FIG{LD}. The longer linkage in fragment 5 (\env{}) might have several reasons that extend beyond our simple argument: (i) homologous recombination might be suppressed in the most variable regions, (ii) the accuracy of \SNP{} frequency estimates is lower in F5 due to poorer amplification, and (iii) the rapid evolution of \env{} due frequent substitutions and sweeps gives less time to break up linkage. In particular, as shown in \FIG{divergenceConsPop}C, frequent and strong selective sweeps affect synonymous diversity in physical proximity along the genome, confirming the presence of linkage at short distances.

\begin{figure}
\begin{center}
\includegraphics[width=\figurewide\figurewidth]{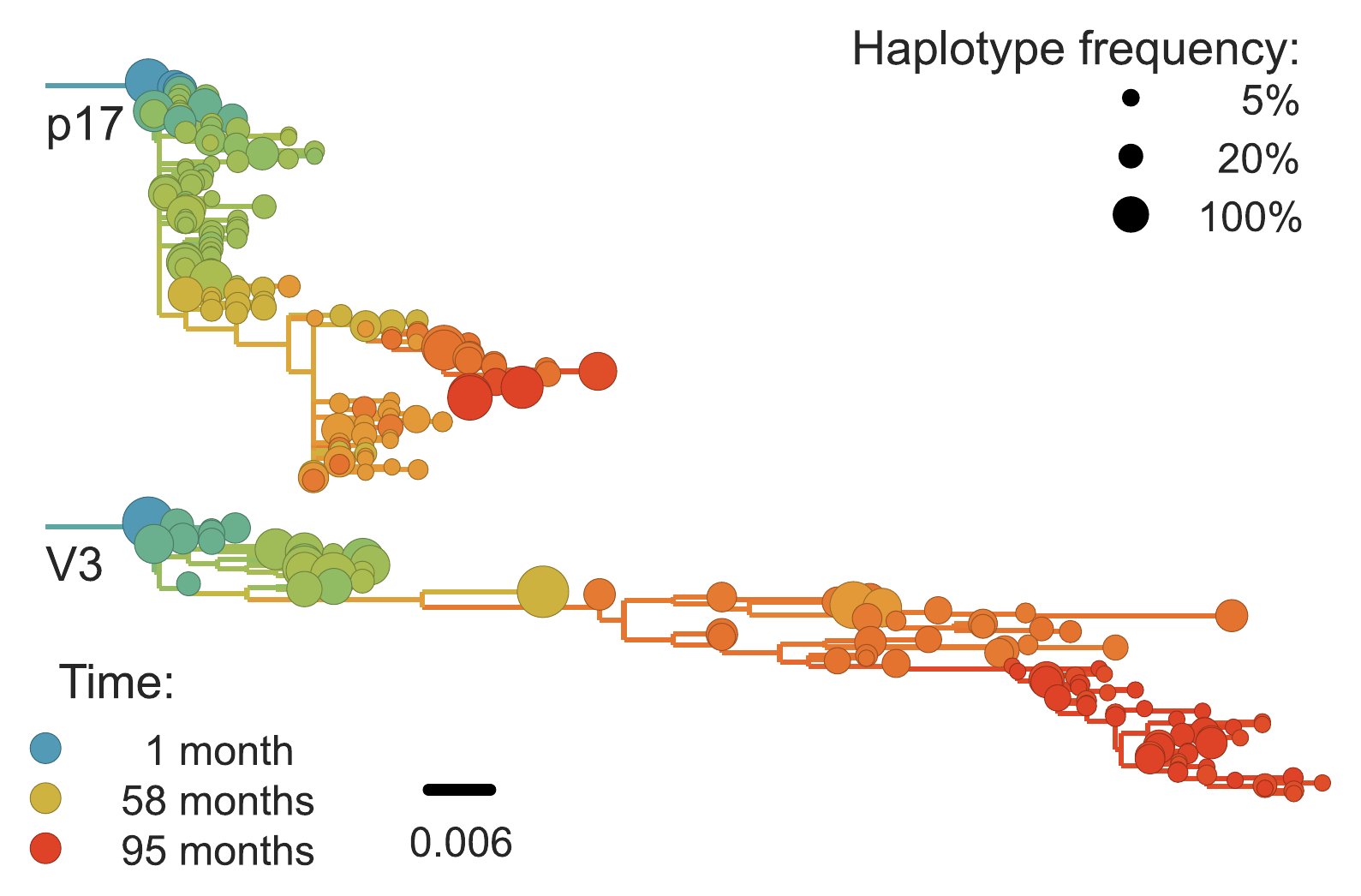}
\caption{{\bf Phylogenetic trees of minor genetic variants. } In rapidly evolving genomic regions, trees that include minor genetic variants (haplotypes) approximate the true phylogeny. Here p17 in \gag{} and the variable loop 3 in \env{} from patient p1 are compared. 
Trees are reconstructed using FastTree \cite{price_fasttree:_2009}.}
\label{fig:trees}
\end{center}
\end{figure}

For phylogenetic analysis, we can extract haplotypes from the sequencing reads up to 500bp in length. Only in the more diverse regions are 500bp sufficient for well resolved phylogenies (see \FIG{trees}). However, we find that linkage does not extend beyond 100-200bp. Hence the read length is not a limiting factor. Only during rapid population shifts such as drug resistance evolution, long read technologies such as PacBio would be necessary to capture the evolutionary dynamics \cite{nijhuis_stochastic_1998}.

\section*{\large{Discussion}} \label{sec:dis}
We have presented a comprehensive portrait of intrapatient evolution of HIV-1 that covers almost the entire genome of the virus, characterizes minor genetic variants, and tracks the fate and dynamics of these variants over a follow-up period of up to 8 years in 9 patients. We find that during the infection HIV-1 explores the sequence space surrounding the founder virus systematically; similar mutational patterns are observed within different, unrelated patients. Linkage between mutations is limited to approximately 100bp, so the virus population can accumulate substitutions independently in different regions of the genome as suggested by theoretical models \cite{rouzine_evolution_2005,mostowy_role_2011}. Nonetheless, local dynamics of \SNP{}s is often dominated by hitch-hiking between neighboring mutations, resulting in an anticorrelation between nonsynonymous divergence and synonymous diversity. A large fraction of all substitutions are reversions towards the global HIV-1 consensus sequence, and these reversions steadily accumulate throughout infection.

The evolutionary dynamics of HIV-1 populations is the result of stochastic forces like mutation and frequent bottlenecks, deterministic fixation of favorable mutations, and recombination. The relative importance of these forces remains unclear \cite{frost_evolution_2000,brown_analysis_1997,rouzine_linkage_1999,kouyos_stochastic_2006,maldarelli_hiv_2013,pennings_loss_2014}. Our observation that intrapatient diversity recapitulates diversity seen across HIV-1 group M and the strong tendency to revert towards consensus suggest that in chronic infection selection determines diversity. The reproducible exploration of sequence space can coexist with frequent adaptation only in frequently recombining large populations \cite{neher_coalescence_2013}. We observe that mutations further apart than 100 bp are effectively shuffled by recombination in most parts of genome, consistent with previous estimates of the HIV-1 recombination rate \cite{neher_recombination_2010,batorsky_estimate_2011}. Linkage and stochastic effects become stronger with increasing frequency of strength of selection, consistent with lower synonymous diversity and more LD in \env{}.

While rapid CTL escape at 5-10 sites over the first two years of infection has been documented in detail \cite{allen_selective_2005,herbeck_human_2006,salazar-gonzalez_genetic_2009,liu_dynamics_2011} and population level associations between specific HLA types and escape variants suggest widespread CTL escape \cite{kawashima_adaptation_2009}, the effect of escape and reversion on long-term evolutionary trends is less clear \cite{lythgoe_new_2012,roberts_structured_2015}. We find a strong tendency for viral populations to revert towards the global HIV-1 consensus. At sites where the founder sequence differs from the subtype consensus, substitutions are almost 5-fold overrepresented: Instead of $\approx 5$\% reversions expected based on the fraction of sequence at which the founder virus differs from consensus of the HIV-1 subtype, almost $25\%$ of substitutions are reversions, in agreement with earlier reports on reversion of CTL escapes \cite{allen_selective_2005,li_rapid_2007}. This tendency to revert increases with the level of conservation of the site, suggesting a quantitative relationship between fitness cost and conservation. While reversion is particularly prevalent in acute infection \cite{li_rapid_2007}, we show that reversion is not limited to early infection but happens throughout chronic infection.

The bias towards reversion results in a two to three fold reduction of the long-term evolutionary rate of HIV, a trend that is reinforced by selection during transmission \cite{sagar_selection_2009,carlson_selection_2014}. Inter-individual evolutionary rates of HIV-1 are two to six times lower than intra-individual rates, and a number of possible mechanisms have been suggested to explain this discrepancy \cite{lythgoe_new_2012}. Our results strongly indicate that most of this mismatch can be explained by steady reversion during infection; other factors such as retrieval of ``stored'' latent variants or stage-specific selection might also contribute to the rate mismatch \cite{lythgoe_new_2012,immonen_reduced_2014}.

The high rate of reversion has implications for phylogenetic dating. Given the 5-fold excess of reverting minor variation, reversion would balance divergence once the typical distance from the consensus sequence equals 17\%, corresponding to a nucleotide diversity of about 30\%; this is remarkably close to the actual divergence between HIV-1 groups M, N and O \cite{li_integrated_2015}. On longer distances, this simple argument will have to be modified due to compensatory mutations resulting in gradual shift of the preferred state at some positions; nonetheless, it indicates a dramatic slowing down of divergence at a scale of the HIV-1-SIVcpz divergence. This apparent deceleration of evolution could explain the contradictory findings of attempts to date the age of HIV-1 and SIV \cite{worobey_island_2010}. The strong and lasting preference for specific nucleotides needs to be accounted for in phylogenetic analysis, as has recently been shown using experimentally determined fitness landscapes of influenza virus proteins \cite{bloom_experimentally_2014}.

The concordance between intrahost variation and patterns of conservation across HIV-1 group M hints at universal fitness costs of mutations. Recently, cross-sectional conservation has been used as a proxy for fitness costs in models of HIV-1 fitness landscapes \cite{ferguson_translating_2013}. Since reproducible intrapatient diversity likely reflects fitness costs of mutations in vivo, our results provide a direct justification for this approach. However, 9 patients are insufficient to extend this analysis to fitness interactions between mutations.

One limitation of this study was the availability of samples from patients with sufficiently long follow-up without therapy after a well-defined time of infection. The majority of patients were MSM infected with subtype B virus. Thus, we cannot exclude that the aspects of HIV-1 evolution that we have investigated differ between transmission routes or HIV-1 subtypes. While substitution and recombination errors of our optimized protocol for HIV-1 RNA extraction and an RT PCR are low, the other main limitation was the at times small number of available template molecules that we quantified by limiting dilution. In principle, the Primer ID method, which labels and resequences each individual template, allows quantification of templates and almost complete elimination of experimental substitution and recombination errors \cite{jabara_accurate_2011}. However, we are not aware of a Primer ID protocol for genome-wide sequencing which was essential to our study.


HIV-1 and other microbial populations evolve in a constant struggle between adaptation to a changing environment and maintenance of functionality. Large mutations rates and population sizes generate standing genetic diversity that is limited by the fitness costs rather than mutation rates. Hence the limiting factor for adaptation is not generating the useful mutations, but combining multiple mutations necessary to survive -- e.g.~escape mutations and reversions -- and pruning deleterious mutations. In HIV-1, this process is facilitated by frequent recombination.

We expect that the systematic exploration of sequence space, the reproducible patterns of minor variation, and frequent reversion will be characteristic of other RNA viruses. Properties of linkage between mutation will differ since mechanisms of recombination are diverse.
But even though selective forces, recombination, and time scales will vary among different microbial populations, theoretical models of rapid adaptation population have shown that many features of the evolutionary dynamics are independent of the specific system  \cite{neher_genetic_2013,fisher_asexual_2013}. Intrapatient evolution of HIV-1 is a unique opportunity to study this evolutionary dynamics directly \emph{in vivo}.

\section*{\large{Materials and Methods}} \label{sec:methods}
\subsection*{Ethical statement}
The study was carried out according to the Declaration of Helsinki. Ethical approval was granted by the Regional Ethical Review board in Stockholm, Sweden (Dnr 2012/505-31/12). Patients participating in the study gave written and oral informed consent to participate.

\subsection*{Patient selection and samples}
The inclusion criteria for the study patients were: a) a relatively well-defined time of infection based on a negative HIV antibody test less than two years before a first positive test or a laboratory documented primary HIV infection; b) no antiretroviral therapy (ART) for a minimum of approximately five years following diagnosis; and c) availability of biobank plasma samples covering this time period. An additional inclusion criterion was used to allow inclusion in an ongoing substudy; successful ART (plasma viral levels <50 copies/µl) for a minimum of two years after at least five years without therapy. The patients were selected from the Venhälsan HIV clinic in Stockholm, Sweden and were diagnosed between 1990 and 2003. Seven to twelve plasma samples per patient (200 µl - 1000 µl) were retrieved from the biobanks of the Karolinska University Hospital and the Public Health Agency of Sweden. These samples had been stored at -70\C\ following routine HIV RNA quantification. Information about the patients and the samples are summarized in Table S1 (\texttt{S1\_samples.csv}). Results from routine HIV antibody tests, HIV antigen tests, plasma HIV RNA levels and CD4 counts were collected from the patient records. Four-digit HLA typing of HLA class I loci A, B, C and class II loci DR and DQ was performed at the Laboratory of Immunpathology, T\"ubingen University Hospital.

In the control experiments we used HIV DNA from the following plasmids; NL4-3 (subtype B, DNA concentration 110 ng/µl corresponding to $1.35 \times 10^{10}$ copies/µl), SF162 (subtype B, DNA concentration 117 ng/µl corresponding to $1.43 \times 10^{10}$ copies/µl), pZM246F\_10 (subtype C, DNA concentration 101 ng/µl corresponding to $1.35 \times 10^9$ copies/µl).

We also used HIV RNA from the following virus isolates LAI III (subtype B, 7,500 copies/µl), 38540 (subtype C, 225,000 copies/µl) and 38304 (subtype B, 45,000 copies/µl).

\subsection*{Estimated time since infection (ETI)}
The patients were classified according to the Fiebig staging system for primary and early HIV infection \cite{fiebig_dynamics_2003}. In addition, we performed  BED tests (Aware\rTM\ BED\rTM\ EIA HIV-1 Incidence Test, Calypte Biomedical Corporation, Portland, OR, USA). For each patient the date of infection was estimated using results from laboratory tests according to the following hierarchical scheme.
\begin{enumerate}
\item Fiebig staging \cite{fiebig_dynamics_2003} was used if results on HIV RNA, antigen, EIA and Western blot were available and the patient found to be in Fiebig stage I-V. Fiebig stages I-VI were considered to correspond to 13, 18, 22, 27, 65 and >100 days since infection based on \citet{cohen_acute_2011}. This was applicable to patients no.~1, 2, 6, 8 and 10.

\item For patients in Fiebig stage VI or with unclear Fiebig stage, ETI was considered to be midpoint between the dates for last negative and the first positive HIV tests if the time interval between these tests was <1 year. This was applicable to
patient no.~5.

\item For remaining patients ETI was calculated using a published time-continuous model of development of antibodies reactive in the BED assay \cite{skar_towards_2013}. However, ETI was not considered to be <100 days for patients in Fiebig stage VI. This was applicable to patients no.~3, 9 and 11.
\end{enumerate}

\subsection*{Primer design}
Primers were designed to cover almost the full HIV genome in six overlapping fragments, called fragments F1-F6 as illustrated in \FIG{sketch}. This allowed sequencing of nucleotide positions 571 to 9567 in the HxB2 reference sequence according to the Sequence Locator Tool available at \url{www.hiv.lanl.gov}. Because of the redundancy of the long terminal repeats (LTRs) this means that all genomic regions except positions 482-571 in the R region of the LTRs were sequenced. Primer design was performed using the subtype reference alignment and the PrimerDesign software available at \url{www.hiv.lanl.gov} \cite{brodin_multiple-alignment_2013}. PrimerDesign was used to find candidate forward and reverse primers targeting highly conserved regions of the HIV genome, with similar melting temperatures, and with minimal tendency for hairpin and primer-dimer formation. Candidate primers were manually adjusted if needed, tested and sometimes redesigned. For each genome fragment both outer PCR primers and nested, inner primers were designed; inner primers were only used for template quantification and internal testing purposes. Alternative primer sets were created for genome fragments F3 and F5 because the PCRs with the original primers sometimes were inefficient. For fragment F5 the amplification problem was not completely alleviated despite trials with several different primer pairs. We believe that this might be be due to the extensive secondary structere in the RRE region. The primers are presented in supplementary table S2 (\texttt{S2\_primer\_list.pdf}). All primer positions except the 5' primer of F1 and the 3' primer of F6 are contained in neighboring amplicons and hence sequenced. Of course, the primer part of the reads itself was trimmed after sequencing (see below).

\subsection*{RNA extraction and amplification}
For each sample, 400 µl of plasma (if available) was divided into two 200 µl aliquots. Total RNA was extracted using RNeasy\rTM\ Lipid Tissue Mini Kit (Qiagen Cat.~No.~74804). Each aliquot was eluted twice with 50 µl RNase free water to maximize HIV RNA recovery. The four eluates were pooled giving a total volume of 200 µl of RNA per sample. The RNA was divided into twelve 14 µl aliquots for duplicate one-step RT-PCR with the outer primers for fragments F1 to F6 and Superscript \rTM\ III One-Step RT-PCR with Platinum \rTM\ Taq High Fidelity Enzyme Mix (Invitrogen, Carlsbad, California, US).

Remaining RNA was used for template quantification (see below). The one-step RT-PCR was started with cDNA synthesis at 50\C\ for 30 min and denaturation step at 94\C\ for 2 min followed by 30 PCR cycles of denaturation at 94\C\ for 15 sec, annealing at 50\C\ for 30 sec and extension at 68\C\ for 90 sec and a final extension step at 68\C\ for 5 min. A second nested PCR was used for template quantification and in some of the control experiments. For the second PCR, 2.5 µl of the product from the first PCR was amplified with Platinum Taq High Fidelity. The second PCR consisted of a denaturation step at 94 \C\ for 2 min, followed by 30 PCR cycles with denaturation at 94\C\ for 15 sec, annealing at 50\C\ for 20 sec and extension at 72\C\ for 90 sec and a final extension at 72\C\ for 6 min. Other PCR conditions were also tried during assay development.

After PCR, the duplicate amplicons from each of the six overlapping PCRs were pooled and purified with Illustra™ GFX™ PCR DNA and Gel Band Purification Kit (28-9034-70, VWR) or AGENCOURT™ AMPure™ XP PCR purification kit (A63881, Beckman Coulter AB). Purified amplicons from each sample were quantified with Qubit™ assays (Q32851, Life Technologies) and thereafter diluted and pooled in equimolar concentrations.

\subsection*{DNA library preparation}
The Illumina Nextera XT library preparation protocol and kit were used to produce DNA libraries. The original protocol was optimized for longer reads and amplicon input in the following fashion: (i) Input DNA concentration to tagmentation was increased to 0.3 ng/µl to reduce overtagmentation; (ii) The number of post-tagmentation PCR cycles was raised up to 14 for samples with very low input DNA; (iii) Post-PCR purification was done using Qiagen Qiaquick columns to maximize large-inserts throughput as compared to magnetic-bead based protocols; (iv) Size selection was performed using the SageScience BluePippin system with 1.5\% agarose gel cassettes and internal marker R2, selecting  sizes of 550 to 900 bp (including dual index Nextera adapters, final insert sizes 400 bp to 700 bp); (v) Size-selected eluates were pooled, buffer-exchanged into EB (10 mM Tris-HCl), and reconcentrated to at least 2 nM.

\subsection*{Sequencing}
The Illumina MiSeq instrument with $2\times250$bp or $2\times 300$bp sequencing kits (MS-102-2003/MS-102-3003) was used to sequence the DNA libraries. We performed 26 paired-end sequencing runs. Overall, we obtained around 200 Gbases of output, i.e. 300 Mbases for each PCR amplicon. The median number of reads per amplicon was 80,000 (quartiles 20,000-220,000, max 2 millions).
All read files have been uploaded to ENA with study accession number PRJEB9618.

\subsection*{Read mapping and filtering}
Bases were called from the raw images using Casava 1.8. The reads were analyzed from that point on using a custom pipeline written in Python 2.7 and C++. We favored this pipeline over existing programs because HIV-1 is a diverse species and both coverage and genetic diversity typically fluctuate by many orders of magnitude across the genome. The pipeline works as follows: (i) reads were mapped onto the HIV-1 reference HxB2, using the probabilistic mapper Stampy \cite{lunter_stampy:_2011}; (ii) mapped reads were classified into one of the six overlapping fragments used for RT-PCR (ambiguous and chimeric reads were discarded), and trimmed for PHRED quality above or equal 30 except for one isolated position per read; RT-PCR primers were also trimmed at this step; (iii) a consensus sequence was computed for each fragment in each sample from a subset of the reads, using a chain of overlapping local multiple sequence alignments (each covering around 150 bp); (iv) the reads were re-mapped, this time against their own consensus; (vii) genetic distance from the consensus were computed, and reads with a distance higher than a sample- and fragment-specific threshold were discarded. Each threshold, calculated to exclude even traces of cross-contamination that might have happened during RNA extraction, PCR amplification, or library preparation, was established by plotting the distribution of Hamming distances of reads from the sample consensus, and excluding reads that are further away than the tail of the main peak. Contaminations appeared as a second peak at higher distances, recombinants as a fat tail: both were excluded. Reads were also trimmed for mapping errors at the edges (small indels); (viii) filtered reads were mapped a third time against a patient-specific reference that was as similar as possible to the consensus sequence from the earliest time point; (ix) reads were re-filtered and checked again for cross-contamination. 
The pipeline was equipped with extensive consistency checks for base quality, mapping errors, and contamination, and is based on the open source projects numpy \cite{van_der_walt_numpy_2011}, matplotlib \cite{hunter_matplotlib:_2007}, Biopython \cite{cock_biopython:_2009}, samtools \cite{li_sequence_2009}, pandas \cite{mckinney_pandas:_2011}, and SeqAn \cite{doring_seqan_2008}.

\subsection*{Quality controls}

\paragraph*{PCR and sequencing errors}
We added 1-5\% of PhiX control DNA to all our sequencing runs. Comparison of reads that mapped to PhiX with the PhiX reference sequence revealed that the Phred quality score were a rather reliable indicator of error rates, with saturation around 0.1\%. \SNP{} calls were calculated from the reads using a custom pile-up restricted to reads with a PHRED quality  $>=30$ corresponding to a minimal theoretical error rate of 0.1\% per base. However, in a dedicated set of control experiments, we established that the main source of error is PCR, see \FIG{afEx} in the main text.

We estimated the error rate of the PCR and sequencing pipeline by amplifying and sequencing plasmid clones (NL4-3, SF162, LAI-III). In most of these experiments we used $10^{4}$ template molecules.
The distribution of \SNPlong{}(\SNP{}) frequencies found in the sequencing reads generated from the plasmid were compared to sequencing results from a typical patient sample, see \FIG{afEx}. After quality filtering, PCR and sequencing errors never exceed 0.3\% of reads covering a particular position. Hence \SNP{}s found in more than 0.3\% of all reads likely represent biological variation.
Note that this neither implies that we detect all biological variants at frequencies above 0.3\%, nor does the frequency of a \SNP{} among the reads necessarily reflect the frequency of that \SNP{} in the sample. For many samples the detection limit and accuracy of quantification of minor variants was not limited by the sequencing error, but by the RNA template input and potential PCR bias.

\begin{figure*}
\centering
\includegraphics[width=0.32\linewidth]{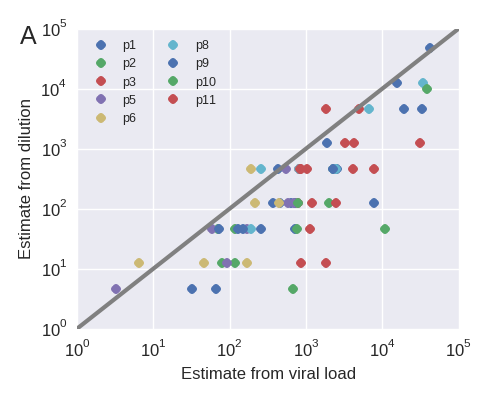}
\includegraphics[width=0.32\linewidth]{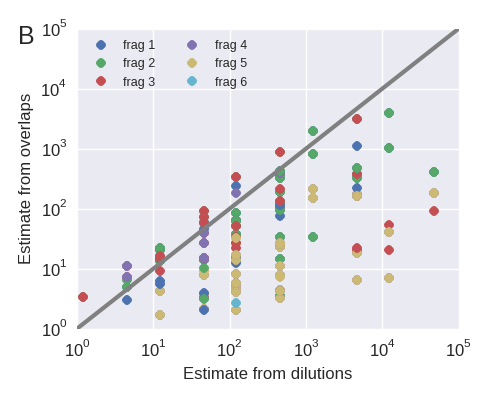}
\includegraphics[width=0.32\linewidth]{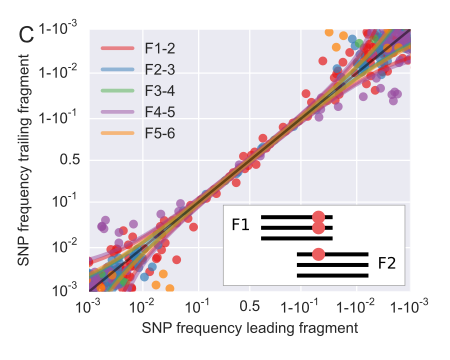}
\caption{{\bf Template quantification and accuracy of \SNP{} frequency estimates.} The left panel shows actual template numbers as experimentally determined by end point dilution ($y$-axis) vs the maximum theoretical template input as estimated based on plasma virus levels. The estimates from the dilutions are typically a factor of three below the expectation from viral load, suggesting generally good template recovery. The center panel show the correlation of fragment specific \emph{effective} template numbers estimated from the concordance of \SNP{}s in fragment overlaps. Again, the correlation is generally good, but fragment F5 often has estimates lower than those from the dilution series consistent with problems faced in amplification of this fragment. Information from fragment F6 is largely absent, since there is very little variation in the F5-F6 overlap. The right panel plots \SNP{} frequencies in overlaps measured in the leading fragment against the trailing fragment. Deviations from the diagonal are due PCR bias and random sampling from limited template molecules. Given enough diversity in overlap regions, the concordance of variant frequencies can be used to estimate fragment specific template input and accuracy of \SNP{} frequencies. }
\label{fig:templateQuant}
\end{figure*}

\paragraph*{Template input}
To roughly quantify the actual number of HIV RNA templates that were subjected to sequencing, we analyzed a dilution series of the RNA (1:10, 1:100, 1:1,000, and 1:10,000) with the nested primers for fragment F4 using the PCR conditions described above. Tests were done in duplicate on the same plate as amplifications for sequencing and the amplicons were visualized by agarose gel electrophoresis. Templates numbers were estimated based on the assumption of Poisson sampling. Thus, for each sample, this dilution series provides an estimate of the number of templates that make it into the sequencing library. The results from the limiting dilution experiments correlated well (rank correlation $\rho=0.7$) with routine plasma HIV-1 RNA level measurements that had been performed at time of sampling, see \FIG{templateQuant}. These comparisons showed that the median template cDNA recovery efficiency was 30\% (quartiles 17\% -- 60\%), which should be regarded as satisfactory since the amplicons were relatively long and the samples had been stored for up to 24 years and sometimes also freeze-thawed prior to this study.

\paragraph*{PCR bias}
In addition to quantification of template input for each sample, we separately estimated the recovery of sequence variation in each of the six sequenced amplicons. Adjacent fragments overlap each other by a few hundred bases. Variation in the overlapping regions is amplified and sequenced on both the leading and trailing fragment completely independently and the concordance of the observed variant frequencies, see \FIG{templateQuant}, can be used to detect PCR amplification bias or poor recovery in a subset of the fragments.

If deviations of \SNP{} frequencies are due to random sampling from the pool of RNA molecules and stochastic PCR efficiency, the variance between the frequencies observed on two neighboring fragments should be $x(1-x)(n_L^{-1}+n_T^{-1})$, where $x$ is the true frequency of the \SNP{}. The numbers $n_L$ and $n_T$ correspond to \emph{effective} template numbers specific to leading ($L$) and trailing ($T$) fragment. For each \SNP{} in the overlap, we calculate the difference $\delta$ and the mean $x$ of the frequencies reported on the $L/T$ fragment. An estimate of $n_L^{-1}+n_T^{-1}$ is then given by
\begin{equation}
\frac{1}{k+p}\left[\sum_i^k \frac{\delta_i^2}{x(1-x)} + \frac{p}{n_{dil}}\right] \ ,
\end{equation}
where $k$ is the number of \SNP{}s in the overlap. To increase robustness of this estimate, we include $p$ pseudo counts each of which contributes $n_{dil}^{-1}$, where $n_{dil}$ is the template number estimated from the dilution series. Since we have 6 fragments but (in the best case) only 5 overlaps, the effective fragment template numbers are not uniquely determined. However, their inverse has to be smaller than the minimum of the estimate of $n_L^{-1}+n_T^{-1}$ at any overlap the fragment participates in. We hence assign a lower bound for the fragment specific template estimate by using the inverse of this minimum, provided with find at least $k=5$ \SNP{}s in the overlap at an average frequency above $0.03$. We use $p=3$. Changing the parameters doesn't change the numbers qualitatively.

The fragment specific estimates of template numbers are consistent with the estimates based on limiting dilution (see \FIG{templateQuant}, rank correlation around $\rho=0.5$ for most of the fragments), but indicated that variant recovery in fragment F5 was sometimes poor, consistent with the difficulties we encountered when amplifying fragment F5. Furthermore, the agreement of \SNP{} frequencies in the overlaps indicate that the primers were well-designed and did not induce significant primer-related PCR amplification biases. These fragment specific estimates are only available for samples and fragments with sufficient diversity in the overlap regions and are often lacking for early samples and fragment F6. The latter is due to the very conserved overlap with fragment F5.

Taken together, these controls show that depending on the sample and fragment, we can estimate frequencies of \SNP{}s down to 1\% accuracy (corresponding to several thousand effective templates). In some cases, however, the template number was low or recovery poor such that no more than presence absence of a variant could be called. All fragment available specific estimates are shown in \FIG{templateQuant}B. In addition to lack of concordance of \SNP{} frequencies in overlaps, problematic fragments lack diversity or have a granular distribution of \SNP{} frequencies such that they can be flagged and removed from specific analysis. The fact that PCR efficiency can vary within one sample is illustrated in Figure 9 -- supplement 1 that compares \SNP{} frequencies estimated in samples 4 and 5 of patient 1 that are only 30 days apart: While SNP frequencies agree to within expectation for 5 out of 6 fragment, fragment F5 shows strong deviations linked to suboptimal PCR efficiency. However, even in problematic samples our amplification and sequencing approach does not generate spurious variation, as shown by the absence of diversity in early samples or conserved sites (see below).

\paragraph*{In vitro recombination and linkage}
The paired-end reads obtained correspond to inserts of up to 700bp in length and therefore provide information on linkage of mutations up to that distance. However, cDNA synthesis and PCR have both the potential to generate in vitro recombination, and true biological linkage is preserved only if the frequency of in vitro recombination is low. To estimate the in vitro recombination in our experimental setup, virions from two subtype B HIV isolates, LAI III and 38304 (the latter obtained from a Swedish HIV-1 patient infected in Brazil), were mixed in equal concentrations. Aliquots of this mix (approximately 1,250 RNA molecules per PCR fragment) were amplified with the six overlapping one-step RT-PCRs as described above in both single PCR and nested PCR mode and then sequenced.

PCR recombination is known to occur predominantly at high amplicon concentrations due to hetero duplex formation of incompletely extended molecules \cite{mild_performance_2011,di_giallonardo_next-generation_2013}). Consistently, we observe no PCR recombination within the first PCR (which starting at low template input does not saturate), while we observe substantial PCR recombination during a second nested PCR. Since our library preparation protocol requires very low input DNA, we do not need nested PCR, avoiding this source of PCR recombination. The fact that PCR recombination occurs during a second PCR shows that the two viruses used for this control (both subtype B) are similar enough that divergence does not interfere with heteroduplex formation. \FIG{LD} includes the linkage disequilibrium observed in the control experiment as a function of distance, showing that linkage is high and not lost with distance during the first PCR.

\subsection*{Analysis}
Python scripts that generate each figure shown in the manuscript are available at \url{github.com/neherlab/HIVEVO_figures}. In these scripts, all parameters, settings, and calculations are explicitly documented.

\paragraph*{Divergence and diversity:} Average divergence defined as the fraction of non-founder alleles in a sliding window of 300bp was regressed against the estimated times since infection using a linear model without intercept. Different data points were weighted by the total divergence, reflecting the expected scaling of the variance if divergence is due to changes at many independent sites. The evolutionary rate was than mapped to the corresponding coordinate in the HxB2 reference.

\paragraph*{Phylogenetic analysis:} To construct trees in different regions of the HIV-1 genome, we extracted reads covering the desired region, restricted them to haplotypes above the desired frequency threshold and constructed a maximum likelihood phylogeny using FastTree \cite{price_fasttree:_2009}. Trees were rooted at the consensus sequence of the earliest available sample.

\paragraph*{Linkage disequilibrium:} To calculate linkage disequilibrium, we first constructed four dimensional matrices for each fragment and each sample, that report the number of times reads mapped such that a nucleotide $n_1$ at position $x_1$ was jointly observed with nucleotide $n_2$ at position $x_2$. 
From these matrices, we calculate the joint frequency $p_{12}(x_1, x_2)$ of the majority nucleotide at positions $(x_1, x_2)$. The normalized linkage disequilibrium measure is then given by
\begin{equation}
D' = \frac{|p_{12}-p_{1}p_{2}|}{D_{max}}
\end{equation}
where $p_1, p_2$ are the frequencies of the majority nucleotide at the respective positions, and
\begin{equation*}
D_{max} = \begin{cases}
\min(p_1p_2, (1-p_1)(1-p_2)) & p_{12}<p_{1}p_{2} \\
\min(p_1(1-p_2), (1-p_1)p_2) & p_{12}>p_{1}p_{2}
\end{cases}
\end{equation*}
Values of $D'$ are then binned by distance and plotted. To reduce noise, only \SNP{}s at frequencies between 20 and 80\% at positions with a coverage $>200$ were included. Furthermore, we required that the fragment specific depth estimate exceeded $40$ or the estimated template input exceeded $200$. The latter is necessary to include fragment F6, since the overlap with fragment 5 often had in-sufficient diversity to allow for a fragment-specific depth estimate.

 \paragraph*{Diversity and divergence}
Synonymous and non-synonymous divergence was analyzed on a per site basis to avoid confounding by different number of transitions and transversions. Positions were classified as synonymous based on the following criterion: The consensus amino acid at this position did not change throughout infection and at least the transition at this site is synonymous. In this case, the observed mutations at this site are almost always synonymous.

The functional categories for genomic regions were the following: structural: p17, p24, p6, p7; enzymes: protease, RT, p15, integrase; accessory: vif, vpu, vpr, nef; envelope: gp120 and gp41.
For the CTL epitope prediction, ranked epitope lists were obtained using the web service MHCi (\url{tools.immuneepitope.org/mhci}). This tool uses several prediction methods ranked based on previously observed performance: Consensus \cite{moutaftsi_consensus_2006} > Artificial Neural Network \cite{nielsen_reliable_2003} > Stabilized Matrix Method \cite{peters_generating_2005} > NetMHCpan \cite{hoof_netmhcpan_2008}.
For each patient, we submitted the consensus amino acid sequences of the viral proteins at the first time point together with 4-digit HLA types at MHC-I.
To analyze putative CTL escape in our patients, we used the first 80 epitopes from the ranked list as a compromise between false negatives and false positives. This number maximized the statistical signature for HIV-1 substitutions being within epitopes, but similar results were obtained by taking the first 50-100 epitopes in the lists.

We preferred computational predictions over experimentally verified CTL epitopes, since the latter are incomplete and biased towards common virus sequences and HLA types. For each patient, we submitted the approximate founder sequence (consensus sequence of the first sample) to MHCi and obtained a list of peptides putatively presented by the HLA alleles (A/B/C) of the respective host ranked by prediction score.

We considered the parts of the HIV-1 genome that putatively is targeted in any of the nine patients, and asked whether the density of nonsynonymous substitutions is higher in the part that is predicted to be presented in a focal patient compared to the part that is not presented. We excluded the variable loops of \protein{gp120} and the external part of \protein{gp41} from this analysis to avoid confounding by antibody selection. Repeating the same analysis for synonymous mutations did not yield any enrichment as expected. The results are insensitive to the number of predicted epitopes used.

\paragraph*{Comparison with subtype diversity}
Full genome HIV-1/SIVcpz sequences were downloaded from LANL HIV data base. Sequences belonging to subtypes B, C, and 01\_AE were individually aligned to HxB2 (using pairwise alignment functions from SeqAn \cite{doring_seqan_2008}) to construct a summary of diversity within subtypes in HxB2 coordinates. Similarly, a group M diversity summary in HxB2 coordinates with made from sequences of subtypes A, B, C, D, F, G, H, restricted to at most 50 sequences from any given subtype. From these HxB2 indexed alignments, entropy was calculated columnwise and the consensus sequence determined as the majority nucleotide.

When comparing HIV variation within patients to cross-sectional diversity, only positions of the founder sequence were used that mapped to HxB2. Similarly, only positions in the cross-sectional alignment were used that had less than 5\% gaps relative to HxB2. This effectively masked regions of the genome that frequently experience insertions of deletions.

To investigate divergence and reversion, intrapatient evolution was studied separately at sites where the approximate founder sequence agreed/disagreed with the respective subtype or group M consensus. Divergence was assessed on a per site basis.

\section{Acknowledgements}
This work was supported by the European research council through grant Stg.~260686. We would like to thank Bianca Regenbogen and Diep Thi Ngoc Tran for help with the design of the web page and Thomas Leitner for helpful feedback and discussion. We would also like to express our gratitude to the study participants.
\bibliography{hiv_data_paper_header}
\end{document}